\documentclass{article}
\usepackage{amsmath} 
\usepackage{graphics} 
\usepackage{graphicx}
\usepackage{xcolor}
\usepackage{bbding}
\usepackage{pifont}
\usepackage{wasysym}
\usepackage{amssymb}
\usepackage[a4paper, total={6in, 9.1in}]{geometry}
\usepackage[labelfont=bf, font=small]{caption}

\usepackage{url}

\begin{document}

\title{\textbf{Detecting periodic time scales in temporal networks}}

\author{Elsa Andres$^1$ \and Alain Barrat$^2$\and Márton Karsai$^{1,3,*}$}

\date{%
\small \emph{
    $^1$Central European University, Quellenstraße 51, 1100 Vienna, Austria\\%
    $^2$Aix Marseille Univ, Universit\'e de Toulon, CNRS, CPT, Turing Center for Living Systems, Marseille, France\\%
    $^3$Alfréd Rényi Institute of Mathematics, Budapest, Reáltanoda utca 13-15, 1053 Hungary\\[2ex]%
    $^*$Corresponding author: mkarsai@ceu.edu
}
}

\maketitle

\begin{abstract}
{Temporal networks are commonly used to represent dynamical complex systems like social networks,  simultaneous firing of neurons, human mobility or public transportation. Their dynamics may evolve on multiple time scales characterising for instance periodic activity patterns or structural changes. The detection of these time scales can be challenging from the direct observation of simple dynamical network properties like the activity of nodes or the density of links. Here we propose two new methods, which rely on already established static representations of temporal networks, namely supra-adjacency matrices and temporal event graphs. We define dissimilarity metrics extracted from these representations and compute their Fourier Transform to effectively identify dominant periodic time scales characterising the original temporal network. We demonstrate our methods using  synthetic and real-world data sets describing various kinds of temporal networks. We find that while in all cases the two methods outperform the reference measures, the supra-adjacency based method identifies more easily periodic changes in network density, while the temporal event graph based method is better suited to detect periodic changes in the group structure of the network. Our methodology may provide insights into different phenomena occurring at multiple time-scales in systems represented by temporal networks.}
{Temporal networks, Fourier transform}
\end{abstract}

\section{Introduction}

Many complex systems, commonly described as networks, are evolving dynamically as their elements and the interactions between them are subject to changes in time. The recent availability of temporally resolved network data sets has stimulated the emergence of the new field of temporal networks~\cite{holme2012temporal, holme2015modern, masuda2016guide}, which has been useful to describe a wide range of phenomena, from human behavior \cite{barrat2013temporal,lehmann2019fundamental,saqr2022and} to biological and ecological systems \cite{trojelsgaard2016ecological,hosseinzadeh2022temporal} or public transportation \cite{huynh2022comparative, salama2022temporal}. The temporal network representation provides an effective tool to investigate the structure and dynamics of these systems, as well as the potential dynamical processes occurring on top of them \cite{barrat2008dynamical,holme2012temporal}. In particular, this representation goes beyond the conventional static description of networks~\cite{braha2009time}, as it keeps track of the temporal order of successive interactions between elements. This allows for instance to identify notions of potential causality through the definition of
temporal paths between nodes, i.e., series of successive interactions along which information can be transmitted~\cite{pan2011path,li2017fundamental}.

Temporal networks present different time-dependent properties at different structural scales. For instance, single nodes can be characterised by their instantaneous degree (number of neighbors at a given time) or other instantaneous centrality measures, which may vary as a function of time
\cite{braha2006centrality,braha2009time,costa2015time,masuda2016guide}. Links are also temporal objects: connections between nodes appear and disappear, 
often following bursty and correlated dynamics, as well as circadian patterns, which are all typical of human dynamics~\cite{barabasi2005origin,karsai2012correlated,aledavood2015digital}.
At mesoscopic scales, temporal networks can exhibit temporal motifs
\cite{bajardi2011dynamical, kovanen2011temporal,longa2022efficient}, communities~\cite{gauvin2014detecting},  core-periphery \cite{csermely2013structure,pedreschi2020dynamic} or other
cohesive structures and hierarchies
\cite{pedreschi2022temporal,galimberti2018mining}.
These various dynamical properties may evolve at different time scales~\cite{saramaki2015seconds}, including overall changes between global states
at the macroscopic level~\cite{gelardi2019detecting,masuda2019detecting,pedreschi2020dynamic}.
In particular, periodic variations can emerge, e.g. driven by the circadian fluctuations of human behaviour~\cite{aledavood2015digital, fournet2014contact}, regular scheduling in different contexts like in transportation or schools, or the repetition of metabolic reactions in biological systems~\cite{lucas2023inferring}. Interesting relevant examples of such variations are given by changes in the connection density in the network, or in the way nodes form and dissolve groups or communities. For example, the number and structure of social interactions vary due to daily rhythms and schedules in contexts such as workplaces, scheduled social gathering or in schools, where students interact within a class
during lectures, but also with other classes during breaks \cite{stehle2011high,fournet2014contact}.
The identification of the temporal scales of periodic variations in a temporal network is an important step for the characterization and understanding of the system under investigation. However, their measure represents a challenge as they co-appear with other arbitrary non-periodic temporal scales, which appear as noise and hinder the possibility to detect the periodic behaviour by simply following the temporal evolution of simple network summary measures.

Some recent works have addressed the detection of relevant temporal scales in temporal networks, e.g. by optimizing the overlap between the sets of events on consecutive time intervals \cite{darst2016detection} or by searching for the precise recurrence of connections between nodes in different time windows \cite{sugishita2021recurrence}. Another approach consists in defining the correlation between instantaneous adjacency matrices of the temporal graph \cite{lacasa2022correlations}.
Finally, computing a whole similarity matrix between all pairs of timestamps
can make it possible to detect states in which the network structure remains
stable
\cite{masuda2019detecting,gelardi2019detecting,pedreschi2020dynamic}, but this method requires rather heavy computations.

Here, we contribute to this endeavour by defining a new method to measure the periodic time scales of temporal networks. Given a temporal network as input, we first divide it into temporal sub-networks using successive sliding windows. We then use lossless mappings of these temporal sub-networks to a sequence of static networks and quantify the dissimilarity between them successively to obtain a dissimilarity function describing the changes between the successive temporal sub-networks. We extract the timescales of this function by taking its Fourier transform to identify its main frequency and harmonics. We focus here on applying this method to the detection of periodic changes in the link density and group structure of temporal networks. To this aim, first we consider synthetic networks in which we impose periodic variations of density and structure with tunable frequencies. We show that the method is able to retrieve the actual time scales of the networks. We then apply our method on several empirical temporal networks presenting periodic dynamics. In each case, the method captures correctly the system's main characteristic times, which could most often not be extracted by simple measures of the network overall activity. Our work opens the door to a better characterisation of the time scales of temporal networks, essential in the understanding of the dynamics of the underlying complex systems.


\section{Methods}

Let us consider a temporal network $G_T=(V, E_T, T)$ defined as a set $V$ of nodes, and a set $E_T$ of events over a time interval $T$ measured in discrete time: each event
$e(i,j,t) \in E_T$ describes a temporal interaction between two nodes $(i,j) \in V \times V$ at a certain time  $t\in T$.

\begin{figure}[!h]
\centering
\includegraphics[width=3in]{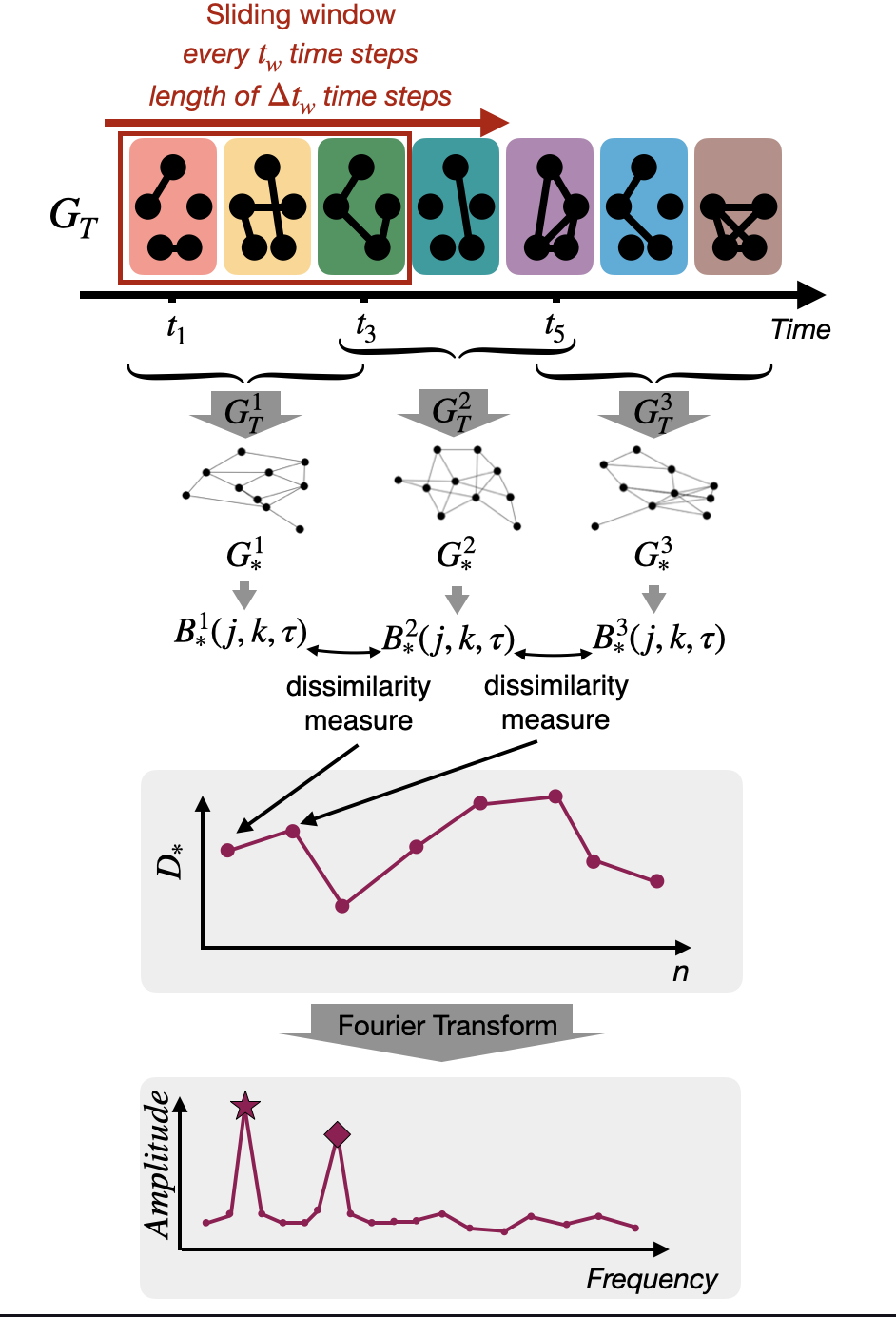} 
\caption{Methodology pipeline to measure the time scales of a temporal network $G_T$. From top to bottom: the initial temporal network is divided into sub-temporal networks through a sliding window. The $m^{th}$ sub-network is denoted $G_T^m$. 
A static representation of each sub-network ($G^m_*$) is generated through the method *. 
Each $G^m_*$ is described by a 3-dimensional tensor
$B_{*}^m(j,k,\tau)$ that encodes information about the paths and distances in the sub-network (see Appendix \ref{appendix:tensor}).
We compare consecutive tensors with a dissimilarity measure, obtaining the dissimilarity function $D_{*}$. Finally, we compute the Fourier transform of $D_{*}$ and measure the frequencies of the main harmonics.}
\label{schema}
\end{figure}

\paragraph{Temporal sub-networks.} With such a temporal network as an input, we first extract a sequence of temporal sub-networks of $G_T$ by using sliding windows of length $\Delta t_w$ and 
stride $t_w$ (shift between the start of successive windows).
Specifically, the $m^{th}$ sub-network $G_T^m$ is composed of the nodes of $V$ and
of the subset $E_{T^m}$ of events of $E_T$ taking place in the time interval starting 
at time $m * t_w$ and ending at $m*t_w + \Delta t_w$:
\begin{equation}
G_T^m=(V, E_{T^m}, T^m=\left[ m * t_w : m * t_w + \Delta t_w \right]) 
\text{\quad for \quad} \, m \in \mathbb{N} 
\text{\quad and \quad} T^m \subseteq T .
\end{equation}
Based on this definition we can obtain a sequence of temporal sub-networks $G_T^m$ to compute a dissimilarity function characterising the dynamical changes in the structure and the overall activities present in the original temporal network $G_T$. More precisely, we want to compute the dissimilarity between consecutive sub-networks,
$G_T^m$ and $G_T^{m+1}$.

\paragraph{Static network representations.} To this aim, we first map each temporal sub-network onto a static network representation using two different methods. Note that both of these representations are lossless and contain the exact same
amount of information as the temporal network they represent:

\begin{itemize}

    \item The \emph{Supra-Adjacency} ($SA$) representation~\cite{valdano2015analytical, sato2019dyane} $G_{SA}=(V_{SA},E_{SA})$ of a temporal network $G_T$ 
    is a static directed network, in which
    each node $v_{SA}\in V_{SA}$ represents a pair (node, time) of the original temporal network: the node $(i,t) \in V_{SA}$ denotes that the node $i \in V$ was active at time $t \in T$, i.e., had at least one interaction at $t$.  
    A directed edge $e_{SA}\in E_{SA}$ between two nodes of $V_{SA}$, $(i, t_a)$ and $(j, t_b)$ (with $t_a <t_b$), encodes the fact that an information can propagate on $G_T$ from node $i$ at $t_a$ to node $j$ at $t_b$, without intermediary events. 
    If $i=j$, this is possible if $t_a$ and $t_b$ are successive interaction times for $i$
 (there is no event involving $i$ at times $t_a < t < t_b$). Edges of type 
 $(i,t_a) \to (i,t_b)$ in $E_{SA}$ thus simply correspond to following the successive interaction times of $i$ in $G_T$.
    For $i \neq j$ instead, the event $(i,j,t_a) \in E_T$ results in two directed edges in $E_{SA}$: $(i,t_a) \to (j,t_b)$ and $(j,t_a) \to (i,t_c)$, where $t_b$ (resp. $t_c$)
    is the first time after $t_a$ in which $j$ (resp. $i$) is active again.
    The direction of edges in $G_{SA}$ respects the arrow of time, and the set of edges $E_{SA}$ allows to preserve the information about all possible temporal paths of the
    original temporal network. 

    \newpage

    \item The \emph{Event-Graph} ($EG$) representation $G_{EG}$~\cite{kivela2018mapping,mellor2019event} is a static weighted directed acyclic network representation of a temporal network. Each event in $G_T$ is represented by a node in $G_{EG}$, and two nodes of $G_{EG}$ are connected if the two corresponding events in $G_T$ were adjacent \cite{kivela2018mapping}, i.e., share at least one node (in $V$) and are consecutive. Each edge between two nodes in $G_{EG}$ is directed along
 the direction of time (from the earlier event to the later one) and is weighted by the
 time difference between the two corresponding events. Consequently, $G_{EG}$ encodes also all information of time respecting paths emerging in the original temporal network.
    
\end{itemize}

These representations can be applied to any temporal network. In particular we apply them to each temporal sub-network $G_T^m$ defined above to map them into a sequence of static representations $G_{SA}^m$ and $G_{EG}^m$. In the following, we use the symbol $*$ to refer to the static representation method: it replaces the abbreviation $SA$ or $EG$, as every object from now on can be calculated using one method or the other.

\paragraph{Network dissimilarity function.} As a next step we compute a dissimilarity function $D_{*}(m)$ between successive static networks, $G_*^m$ and
$G_*^{m+1}$, for each sequence of static representations $\{G_*^m, m=1,\cdots\}$.
To this aim, we consider here an extension of the method of \cite{bagrow2019information}
that summarizes the properties of a network in a matrix where the element $(k,l)$ gives the number of nodes that can reach $k$ other nodes in $l$ hops in the structure. As we originally deal with temporal networks, and as the nodes of the $SA$ and $EG$ representations do keep temporal information, we instead describe each $G_*^m$ by a tensor $B^m_{*}(j, k, \tau)$, which gives the number of nodes in $G_*^m$ from which one can reach, in two hops on $G_*^m$, other nodes of $G_*^m$ involving $j$ nodes, $k$ events and $\tau$ timestamps of $G_T^m$.  The dissimilarity function $D_*^m$ is finally computed as the Kullback-Leibler divergence between $B^m_{*}$ and $B^{m+1}_{*}$.

\paragraph{Fourier Transform of dissimilarity function.} Each dissimilarity function $D_*$ provides an overall signal that reflects the structural and activity changes in the original temporal network. It presents higher values when the network goes through larger and abrupt transformations and takes smaller values when the network is more stable or changing only gradually with time. It can thus provide insights into the time scales of dynamical changes in the original temporal network. In particular, periodic patterns of network changes can be revealed by taking the Fourier transform of the dissimilarity function, which should present harmonics at the characteristic frequencies of the temporal network. More precisely, we compute the discrete-time Fourier transform of $D_{*}$ defined as:
\begin{equation}
FT_k = \left| \sum_{j=0}^{N_{sample}} D_*(j) e^{i2\pi k j / N_{sample}} \right|,
\end{equation}
where $N_{sample}$ is the length of $D_{*}$ and $k \in [0, N_{sample}-1]$. The frequency corresponding to the $k^{th}$ harmonic $FT_k$ is $f_k = \frac{k}{t_w N_{sample}}$ where $t_w$ is the time shift between two successive sub-networks $G_T^m$. The main harmonics of the $FT$ function (appearing as the largest modes in the transformed function) correspond to the principal frequencies of the temporal network. Their inverse yield the characteristic time scales of the main periods present in the network dynamics. 

In the following, we refer to the full methodology pipelines using respectively the $SA$ and $EG$ representations as the $SA-method$ and $EG-method$. The whole generic pipeline is summarized in Figure~\ref{schema}.


\section{Validation on synthetic data sets}

To better understand the temporal properties that the above defined dissimilarity functions and their Fourier transforms can capture, we focus on synthetic temporal networks with controlled structural and temporal properties. In particular, we consider networks 
with tunable changes in activity (number of events per timestamp) and group structure.
We utilise the Activity-Driven temporal network (ADN) model~\cite{perra2012activity} for these purposes, defined by a set of $N$ nodes $i=1,\cdots,N$, each having an intrinsic activity $a_i$
taken from a given distribution. At each time step, node $i$ becomes active with
probability $\eta a_i$ and, if active, establishes connections with $m$ other
nodes chosen randomly. Connections are erased after each time step thus the model does not present any memory nor correlations between time steps. Here we consider networks of size $N=100$ with a power-law
node activity distribution with minimum value $\epsilon = 0.001$ and parameters $\gamma=1.8$, $m=4$, $\eta=4$ and $\mid T \mid=9200$.

\begin{figure}[!h]
\centering
\includegraphics[width=0.9\textwidth, trim={0cm 0cm 0cm 0cm},clip]{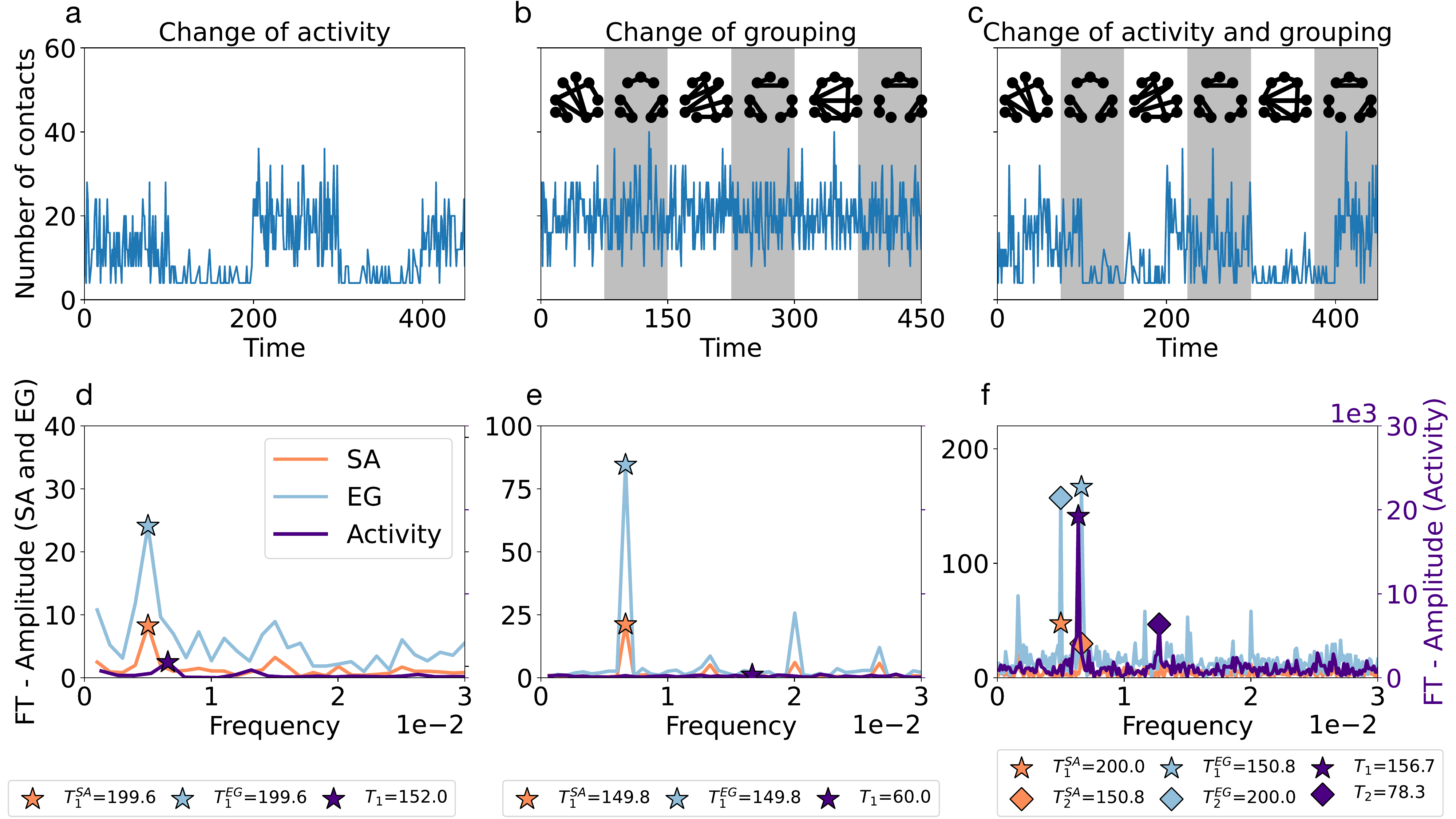}
\caption{
Schematic representation of three settings simulated with the Activity-Driven temporal network model with periodic changes of parameters ($N=100$, $\epsilon=0.001$, $\eta = 4$). (a) The \emph{Change of activity} case presents networks with activity periods of $T_a=200$; (b) the \emph{Change of grouping} case presents recurrent structural changes with period $T_g=150$; while (c) the \emph{Change of activity and grouping} setting is defined as a mix of both dynamics. Panels (a-c) display the number of events as a function of time for a realization of each experiment; Gray areas in panels b and c indicate the intervals in which interactions can only occur within groups. Panels (d-f) depict the Fourier transforms of these networks obtained respectively through the {SA-method} and {EG-method}, as well as the Fourier transform of the activity timeline. The first and second harmonics of each Fourier transform are shown respectively with a star and a diamond symbol. In each case, the {SA-method} and {EG-method} are able to retrieve the correct period of the networks, while the Fourier transform of the activity signal fails in measuring temporal structural changes. In the \emph{Change of activity and grouping} case, the {SA-method} identifies the frequency of activity changes as the main harmonic, while the {EG-method} detects the structural changes frequency as the dominant one.
}
\label{experiements}
\end{figure}

Using these parameters as baseline, we build three types of periodically varying temporal networks, to model the following settings:
\begin{itemize}
    \item \emph{Change of activity}: we simulate an ADN in which the density of edges varies periodically in time. We assign to each node $i$ two activity values $a_i^1$ and $a_i^2$, respectively extracted from two power-law distributions with exponents $\gamma_1=1.8$ and $\gamma_2=2.8$. We then alternate periodically (and synchronously for all nodes) between the two activity values, with a period $T_a$. This results in periodic changes in the overall activity of the network, as illustrated in Figure~\ref{experiements}a. 
   
    \item \emph{Change of grouping}: we consider an ADN model of $N=100$ nodes forming groups of $5$ nodes each, and we periodically alternate, with a period of $T_g$, between time intervals in which connections are made at random with no restriction as in the baseline and intervals in which only connections within groups are allowed. The average activity is kept constant over time (Figure~\ref{experiements}b).
  
    \item \emph{Change of activity and grouping}: finally, we consider an ADN
    in which both activity and group structure change periodically over time, by combining the previous two mechanisms, each with its own period, respectively $T_a$ and $T_g$ (see Figure~\ref{experiements}c).
\end{itemize}

For each case, we apply the SA and EG methods to compute the Fourier transforms of the resulting temporal networks. As a baseline method, we compute directly the Fourier transform of the activity function, that is measured as the link density at each time step of observation (see Figure \ref{experiements}). This is a simple summary metrics that describes the overall changes in the temporal network and can be computed for any system.

\subsection{Results}

The settings we consider involve either one or two types of periodic changes in the synthetic temporal networks: a periodic fluctuation in the amount of activity and/or in the network structure in terms of inter and intra group interactions. Our first goal is to investigate whether the SA- and EG-methods can uncover the corresponding periods $T_a$ and $T_g$ through the measure of the dominant frequencies in the associated Fourier transforms. As shown in Figure~\ref{experiements}d and e, when only one type of periodic change is present, both methods are able to detect the corresponding period. It is evident from the depicted star symbols that indicate the largest mode in the frequency scale, correctly positioned at the right frequency corresponding to the period of the actual periodic changes. At the same time, the baseline method, computed as the FT of the activity timeline, strongly underperforms as compared to the other two methods. While in case of activity changes (see panel Figure~\ref{experiements}d) it at least identifies approximately the value of the period, in case of periodical group changes it does not succeed to capture the rightful period at all. This was expected as in this case the overall activity does not reflect any periodicity but simply fluctuates randomly around a constant value.

When both types of periodic changes are present, an interesting distinction emerges between the results of the SA- and EG-methods. Indeed, both methods correctly detect the $T_a$ and $T_g$ periods as the first two dominant frequencies in the Fourier transform. However, in the {SA-method} the frequency describing the periodic activity changes is identified as the dominant frequency
and the periodic group frequency to the second largest value
($T_1^{SA}=T_a=200$, $T_2^{SA}=T_g=150$ in Figure~\ref{experiements}f), 
while this is reversed for the EG-method
($T_1^{EG}=T_g=150$, $T_2^{EG}=T_a=200$ in Figure~\ref{experiements}f).
These results suggest that the {SA-method} is more sensitive to periodic changes in activity, while the {EG-method} is more suited to detect periodic structural fluctuations. We also note that the FT of the baseline method yields as dominant timescales $T_1=156.7$ and $T_2=82.5$, the first one describing approximately the activity periods of the network, while the second one does not correspond to the period of either of the underlying processes.

\begin{figure}[!h]
\centering
\includegraphics[width=0.95\textwidth, trim={0cm 5cm 0cm 3.5cm},clip]{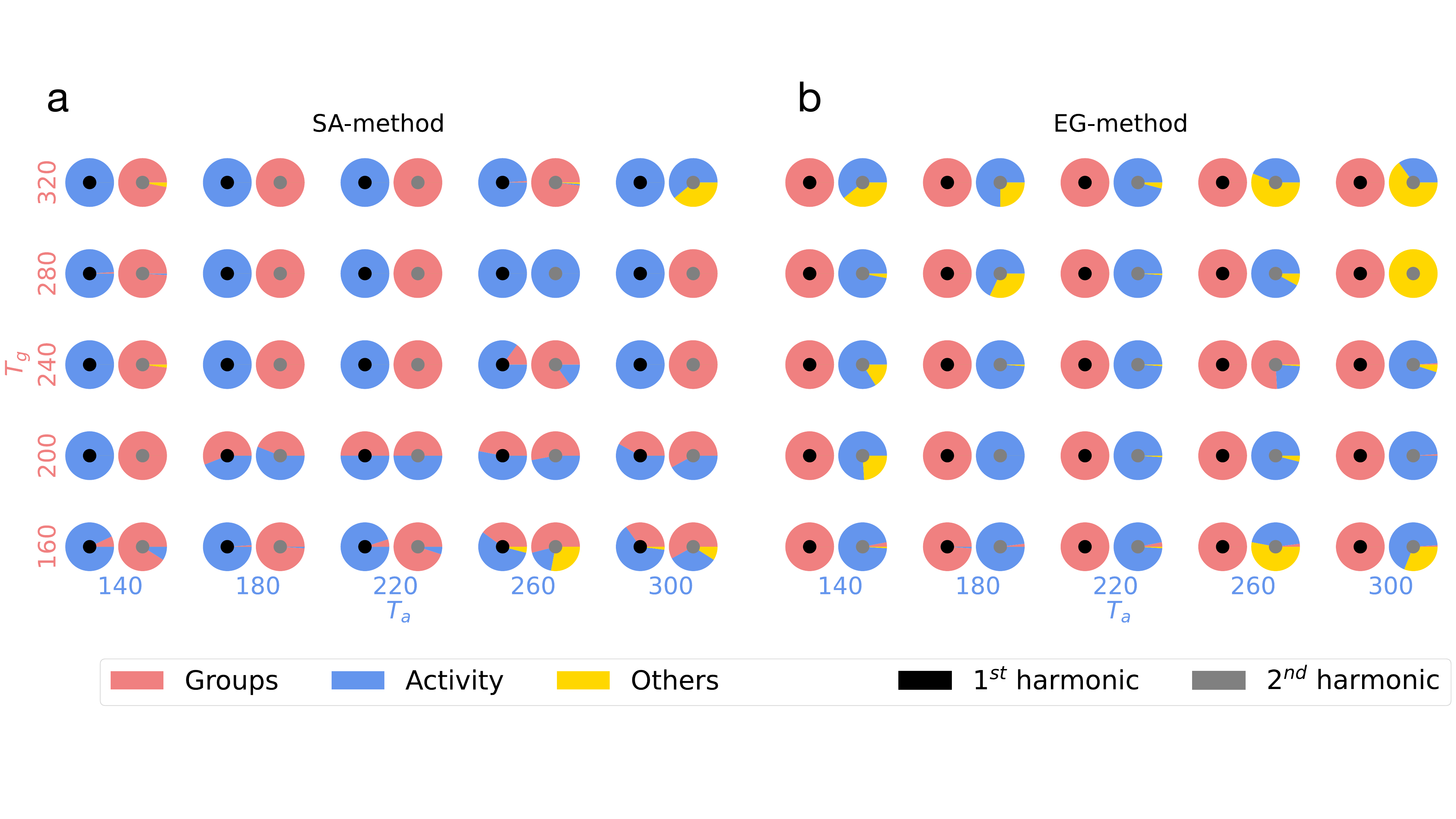} 
\caption{Periods corresponding to the two first harmonics measured through the {SA-method} (panel a) and the {EG-method} (panel b), for periodic synthetic temporal networks generated through the \emph{Change of activity and grouping} setting 
($N=100$, $\epsilon=0.001$, $\eta = 3$, $\mid T \mid=9200$) with respective periods $T_a$ (x-axis)
and $T_g$ (y-axis).
For each pair of values ($T_a$, $T_g$), we generate $100$ realizations of the temporal
network and apply the SA- and EG-method to extract the two main harmonics. 
We show in blue around a small black disk (resp. grey disk)
the fraction of realizations in which the main frequency (resp. the second main)
corresponds to $T_a$, in pink the fraction of cases in which it yields 
$T_g$, and in yellow the cases in which it corresponds to neither
(we consider a tolerance of $10\%$ for both periods).
In most cases, both periods are correctly inferred, with the main frequency corresponding to $T_a$ in the SA-method and to $T_g$ in the EG-method.
}
\label{fig:mixing}
\end{figure}

To check the robustness of the proposed methods against the relative values of the periods, we further investigate this point by exploring systems with different values of $T_a$ and $T_g$ in the \emph{Change of activity and grouping} setting. We generate $100$ synthetic temporal networks for each pair of values ($T_a$, $T_g$), compute the dissimilarity function and Fourier Transforms of these realization, and extract the corresponding first two harmonics for each method (SA and EG).

Figure \ref{fig:mixing} summarizes the results by showing in each case the fraction of
realizations which detected the periods of $T_a$, $T_g$ correctly, or failed to detect any of them. These results demonstrate again that the SA-method identifies predominantly $T_a$ (the activity change period) through the first harmonic and $T_g$ (change of group structure) through the second, while the reverse
is observed for the EG-method. Some deviations from this behaviour are observed at large
values of the periods and/or when $T_a$ and $T_g$ are close to each other.

\subsection{Parameter dependencies and limitations}

Both the synthetic temporal networks and the analysis method involve some parameters. In particular, we explore their dependencies on the network size, temporal length and ratio between total temporal length
and periods of changes $T_a$ or $T_g$, with results presented in Appendix \ref{appendixa}. We observe that at large network size, both methods identify as main frequency a value corresponding the half value of the original period. Moreover, evidently, for correct time scale detection the observation period of the temporal network need to cover at least two full periods of any kind of changes.

The first step of our pipeline moreover involves the definition of sliding windows with stride $t_w$ and length $\Delta t_w$. Naturally, these parameters affect the amount of information contained in each sub-temporal network and consequently influence the resulting dissimilarity function~\cite{sulo2010meaningful,krings2012effects,kivela2015estimating}.
We explore the effect of these parameters in Appendix \ref{appendixa}, while keeping $t_w \le \Delta t_w$ to have a non-zero overlap between successive time windows. We also ensure that the two parameters under study have values below the time span of the network's period (their maximum value is $20$ while the period is $100$).


As shown in Appendix B in Figure~\ref{appendix1b}, the two methods show the best performance if the $t_w$ stride is not too large and if $\Delta t_w$ length is neither too high nor too small. 
If the time interval between two temporal sub-networks $t_w$ is too high, we collect less information about the similarity between successive sliding windows. The dissimilarity function is then less precise and our methods perform less well to identify the characteristic temporal scales. Moreover, if $\Delta t_w$ is too small, each sliding window contains too little information to obtain an accurate measure of the time-scale of the original network.
On the opposite, if $\Delta t_w$ is too large, each temporal sub-network may summarize too much information and loose the specific characteristic of the activity or the structure of the network on a certain time or interval of time. As an observation bias this could smooth dissimilarities between consecutive temporal network slices as they average too much information, and not because the network does not present significant changes through time. 

It is also worth noticing that both methods measure systematically half of the period as dominant modes for very large values of $t_w$ and $\Delta t_w$ (Appendix B, 
Figure~\ref{appendix1b}). 
In that case, every half-period of the network is covered by a small number of temporal sub-networks, leading to a lack of resolution in 
the dissimilarity function, in which only the peaks of dissimilarity at half-periods are well marked, leading to the detection of the half-period as typical timescale.


\section{Applications on real networks}

After validating our methods on synthetic networks with controlled properties, to explore further the capabilities of our methods, we consider empirical temporal networks representing different systems. We note that in such systems, in contrast to the cases studied above, several time scales, that correspond both to periodic or non-periodic fluctuations, may co-exist, as well as structural changes of different nature.

\subsection{Data sets}

We consider four temporal networks describing interactions of different nature, with various sizes and over different observation lengths. For more details about their temporal dynamics see Appendix \ref{appendixc}.
\begin{itemize}
    \item \emph{US middle school network}: this data set describes close proximity interactions between students of a middle school in the United States, during one day with temporal resolution of $20$ seconds \cite{toth2015role}, recorded by Radio Frequency Indentification (RFID) wearable devices. It involves several periods of class-times and inter-class breaks including two lunch periods, when students freely mix while changing classroom or eating together. The network consists of $591$ nodes (each node corresponding to a student) and contains $473,755$ records of pairwise temporal interactions between them.

    \item \emph{Conference network}: these data also describe face-to-face contacts between individuals, with a temporal resolution of $20$ seconds, obtained by RFID devices built on a different architecture \cite{cattuto2010dynamics}. The contacts were measured during a scientific conference, namely the IC2S2 conference that took place in Cologne (Germany) in 2017 \cite{genois2023combining}. Our observation period spans over the three first days of the conference, and records $229,536$ temporal contacts between $274$ participants. This data set, similar to the school data, is expected to show periodic behaviour both in terms of activity and structural changes, by reflecting the scheduled sessions and session-breaks of the conference.

    \item \emph{Resistance game network}: it is an eye-contact network between participants of the Resistance game \cite{bai2019predicting, kumar2021deception}, which is a role game where some of the players are hidden 'defeaters', and the goal of the other players is to uncover them. The game involves multiple rounds of around $4$ minutes each, starting with a discussion involving every participant, and ending with a vote. The recorded network is built from directed events between participants who looked at each other at a given time $t$. The network is recorded between $8$ individuals and contains $52,731$ temporal interactions that we deem undirected for simplicity. This network provides an example where the interaction level should not reflect strong periodicity but the grouping of participants changes between each session.

    \item \emph{US flight network}: this air-transportation network describes the direct flight connections between $278$ airports in the US \cite{airplane}. In our observation period we concentrate on $4$ days of data that records $71,315$ flights between the airports that we consider as undirected temporal interactions. This network is expected to show strong periodicities in activity, reflecting the daily recurrent flight schedules, while structural changes may not be strong as almost always the same airports are connected every day.

\end{itemize}

\subsection{Results}
\label{results_real_networks}

\begin{figure}[h!]
\centering
\includegraphics[width=6in]{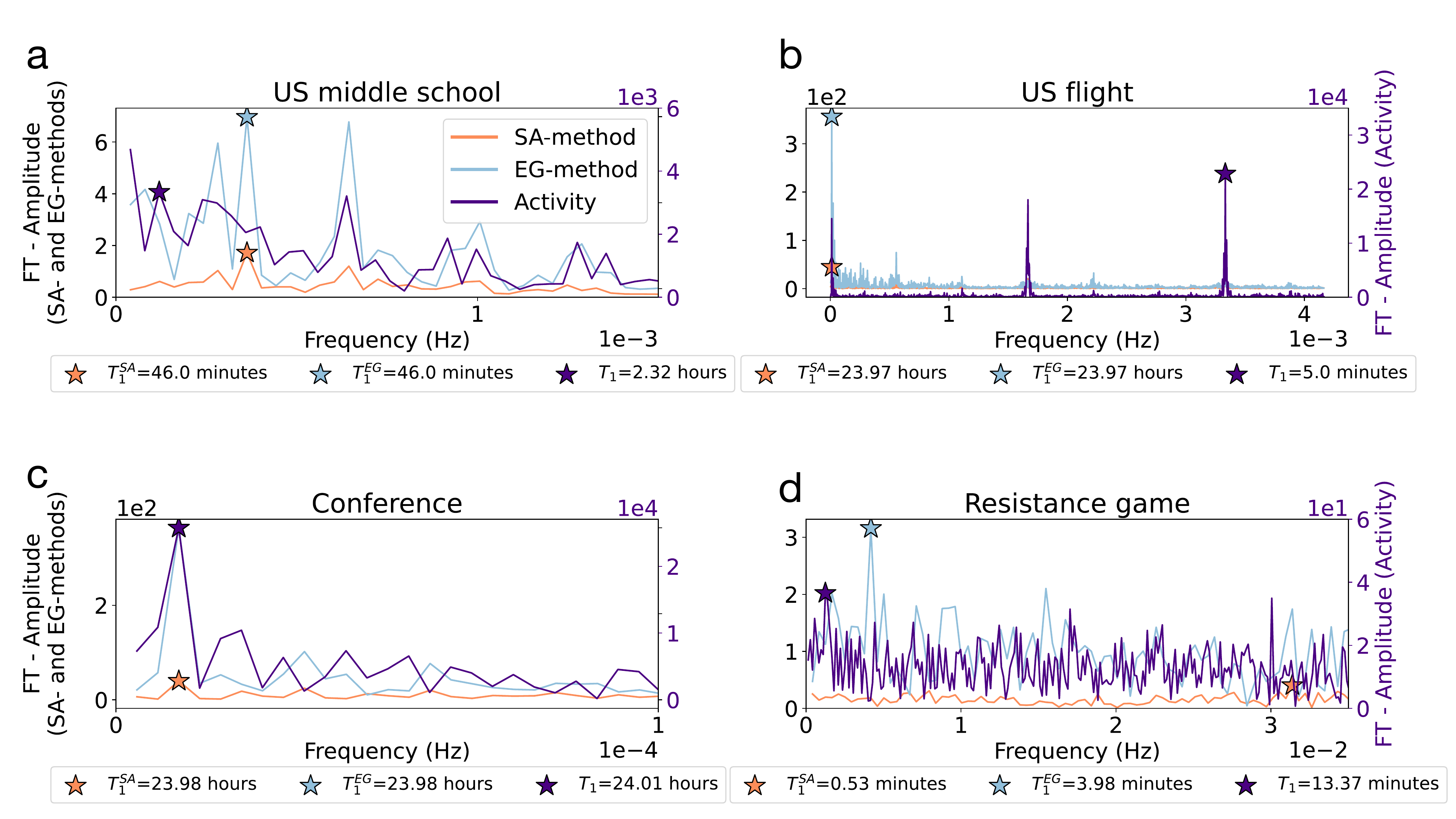}
\caption{
Fourier transforms of dissimilarity and activity functions of four real-world data sets (a) a US middle-school, (b) the US flight network, (c) a conference, and (d) the resistance game networks. Dissimilarity functions were calculated by the {SA-method} (in orange) and the {EG-method} (in blue), while results computed for the baseline model using activity signals are shown in purple. The highest harmonics are highlighted with a star symbol for each FT, and the corresponding values of the period is indicated below each panel. The parameters of the sliding windows $(t_w, \Delta t_w)$ are (2 minutes, 5 minutes) for the \emph{US middle school}, (1/3 minute, 1 minute) for the \emph{Resistance game}, (2 minutes, 10 minutes) for the \emph{US flight} and (2 minutes, 5 minutes) for the \emph{Conference}. 
}
\label{real_data}
\end{figure}

After applying our pipeline on each data set using both the SA- and EG- and the baseline methods, in Figure \ref{real_data} we depict the Fourier transforms of the obtained dissimilarity functions, with stars indicating the dominant frequencies. Interestingly, both the SA- and EG-methods identify the relevant timescales in most networks, while the baseline method consistently failed to detect them. For the \emph{US middle school network}, both methods yield a timescale of about $46$ minutes, coherent with the length of a class. Meanwhile, the baseline activity timeline FT would estimate the dominant frequency as corresponding to a period of $139$ minutes. In the case of the \emph{US flight network}, where the main changes are expected to be ruled by circadian fluctuations, both SA- and EG-methods also correctly identified periods of around $24$ hours. This time-scale is also captured by the baseline method, but recognised only as its third largest harmonic. The two first harmonics are identified as periods of $5$ and $10$ minutes, which may correspond to the characteristic times between consecutive departures of planes from the same airport. The \emph{Conference network} also presents strong signs of circadian changes of activity. This is reflected by all computed Fourier transforms, which show a harmonic corresponding to a period of about 24 hours for both the {SA-method} and the {EG-method}, captured as well by the baseline method. Finally, regarding the \emph{Resistance game}, which presents only structural changes, the {EG-method} measures accurately the time-scales of periods characterising a single round in the game, around 4 minutes. Since no periodic change of activities characterise this network, both the SA-method (more sensitive to activity changes) and the baseline method fail to identify any meaningful time-scale. The Fourier transform of the SA-method suggests the dominant mode to correspond to $0.53$ minute, while the baseline method detects $13$ minutes.

\subsection{Shuffling of the data}

Empirical temporal network data entail structural and temporal correlations of different nature. To explore which of their characteristics play the main role in determining their relevant time-scales, a common method consists in shuffling the data to create randomized
reference models \cite{gauvin2018randomized} in which specific correlations are destroyed while others are preserved. In other words, through shuffling we create a sample from a uniformly sampled microcanonical ensemble of randomized networks, where certain network properties are kept constrained, while the networks are maximally random otherwise. Specifically, here we consider shuffling methods that remove the periodicity of the activity and/or of the group structure. In turn, we apply the SA- and EG-methods to compute the Fourier transforms of the shuffled data and check how these methods capture the modification of the time scales due to shuffling. We consider the following random reference models, following the canonical notations introduced in~\cite{gauvin2018randomized}:

\begin{itemize}
    \item \emph{$P_p(\Gamma)$} shuffling: To remove both activity and structural correlations, we randomly shuffle the order of the temporal network snapshots, keeping fixed the structure of each snapshot. This procedure destroys any structural correlations between consecutive snapshots, removing the effects of structural reorganizations and randomizing also the activity timeline.
    
    \item \emph{$P_\tau$} shuffling: we rewire randomly all links in each snapshot of the temporal network. This is equivalent to creating a configuration random network~\cite{newman2018networks} in each snapshot, with the same number of nodes and edges as the original snapshot. The activity timeline is thus preserved while the group structure and its changes are removed.
\end{itemize}

\begin{figure}[h!]
\centering
\includegraphics[width=0.85\textwidth, trim={0cm 0cm 31cm 0cm},clip]{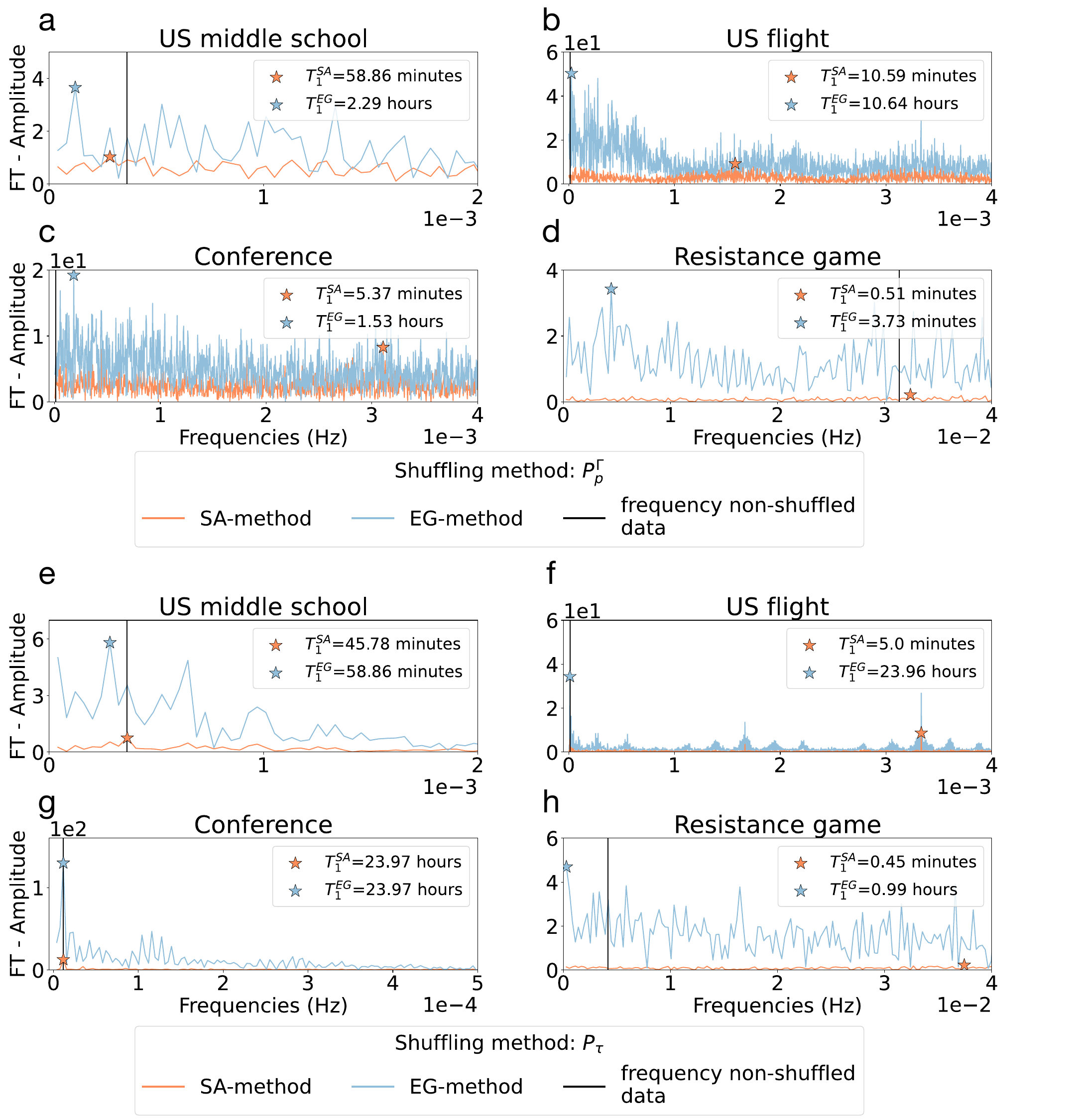}
\caption{
Fourier transform for the data sets {US school}, {Resistance game}, {US flight} and {Conference} networks shuffled using 
 the two shuffling methods $P_p(\Gamma)$ (panels a-d) and $P_t$ (panels e-h), obtained with the {SA-method} (orange curve) and the {EG-method} (blue curve). The period of each original data set is indicated with a black vertical line. 
 For data shuffled using the $P_p(\Gamma)$  method, the original period is never recovered.
 In the case of the $P_t$ shuffling instead, the {SA-} and {EG methods} still measure  original periods if the network presents large activity changes (\emph{US flight} and \emph{Conference} data sets).
 In the case of the \emph{US middle school} network, only the {SA-method} is able to assess the original time scale as this method performs better to detect activity changes. Finally, none of the method can measure the original period of the \emph{Resistance game} network shuffled with the $P_t$ method as it does not present any periodic variations.
}
\label{fig_shuffle}
\end{figure}

Results are shown in Figure \ref{fig_shuffle} for the four data sets and the two shuffling procedures. When the networks are shuffled with the $P_p(\Gamma)$ procedure (panels a-d), the original periods are not recovered, which is expected since the shuffling destroys any periodicity in the data.

However, when we shuffle the networks using the \emph{$P_\tau$} method, which removes the structural effects but keeps the fluctuations in the overall activity, our methods present some capacity to identify the residual time scales in some of the data sets. 
In particular, two of the data sets 
present large periodic activity variations, i.e. the \emph{US flight} and the \emph{Conference} networks. After shuffling, these regular changes
are still present, as the \emph{$P_\tau$} method preserves their activity time line, while any other pattern has been destroyed by the shuffling. Consequently, we may still measure their original circadian period from their \emph{$P_\tau$}-shuffled versions. Indeed, both the {SA-} and the {EG-methods} applied to the \emph{Conference} network recover the dominant time scales, while in case of the \emph{US flight} data set, the {EG-method} captures the expected time scale of around one day. We also find a time scale of 5 minutes with the {SA-method} applied to the \emph{US flight} data set, which corresponds to another characteristic times of activity of this network 
(see Figure \ref{real_data}).

In contrast, both the {SA-} and the {EG-methods} miss the identification of the original time scales when applied on the \emph{$P_\tau$}-shuffled \emph{Resistance game}. Since the original network has no periodic fluctuations in terms of activity, neither its shuffled counterpart present any regular changes in term of activity. Thus the detected time scales are only induced by some noise in the data.

Finally, the \emph{US middle school} network presents activity variations that are not easily  assessed even in the original network. Once shuffling with the \emph{$P_\tau$} method, only the {SA-method}, which is overall more sensitive to activity changes, retrieves the original period 
($\approx 46$ minutes) in the shuffled network. The {EG-method} overestimates this time by detecting a period
of $5\approx 9$ minutes.

\section{Conclusion}

In this work, we have put forward a new methodology to uncover periodic time-scales in temporal networks. In our proposed pipeline, first we locally aggregate the original temporal network by using a sliding window to build a sequence of temporal sub-networks. Subsequently, we map these temporal sub-networks into a sequence of static networks, using known lossless higher-order temporal network representations, namely supra-adjacency matrices or event-graphs. 
We further extend a method for the comparison between the consecutive static network samples to define a dissimilarity function that reflects activity and structural changes in the original temporal network. Finally, we  take the Fourier transform of the dissimilarity function to detect the relevant periodic time-scales from the dominant frequencies characterising the original network.

We have explored this pipeline, focusing on changes in the activity and group structure
of temporal networks. Using synthetic data sets with prescribed changes, we have shown
that while both methods are able to recover the time scales of the modelled periodic dynamics, they perform differently in the identification of changes in activity and structure. Specifically, the SA-method is more sensitive to overall activity changes while the EG-method captures better periodic structural fluctuations, which cannot instead be obtained through the FT of the activity timeline. We have also shown that these methods are able to highlight relevant periods in more complex empirical data sets.

The methodology presented here have certain limitations. First, its performance depends on some parameters of the aggregation method and the temporal network observed. The observation needs in particular to span a long enough interval: at least two periods of changes need to be observed. The sliding window parameters also
have some impact on the performance: each temporal sub-network should encode enough information but should not be too long to average out relevant changes. The stride should be small enough to keep a reasonable temporal resolution and a substantial overlap between successive windows.

The proposed methodology pipeline opens the door to the investigation of several interesting extensions and research questions. Possible extensions of the present method could include the consideration of other static representations as well as other similarity measures between successive 
temporal sub-networks \footnote{including the tensor portrait defined in Appendix \ref{appendix:tensor} but considering different numbers of hops from each node.}, which could potentially be more sensitive to various types of structural changes of the temporal network. For instance, 
it would be interesting to explore whether changes in the instantaneous core-periphery structure \cite{pedreschi2020dynamic} could be uncovered. Future work could also explore extensions to time-varying hypergraphs~\cite{battiston2020networks,battiston2021physics} or the interaction between the detected time scales of the underlying temporal network and ongoing dynamical processes.
Our work presents a proof of concept for a new methodological direction that will contribute to the better characterisation of time varying complex structures. 

\section*{Acknowledgment}

We acknowledge support from the Agence Nationale de la Recherche (ANR) project DATAREDUX (ANR-19-CE46-0008). MK was supported by the CHIST-ERA project SAI: FWF I 5205-N; the SoBigData++ H2020-871042; the EMOMAP CIVICA projects and the National Laboratory for Health Security, Alfréd Rényi Institute, RRF-2.3.1-21-2022-00006.

\appendix 

\section{Comparing temporal networks}
\label{appendix:tensor}

An important step in our methodology is to quantify the similarity between successive temporal sub-networks. We first map each temporal sub-network to a static one, as described in the main text, and compare these static representations.
To this aim, we adapt a method proposed by Bagrow and Bollt 
\cite{bagrow2019information}, which allows to compare static networks at multiple scales. The first step of this method is to compute, for each static network, its ``portrait'' $B$ defined as 
\[
B_{l,k} = \text{number of nodes which have $k$ nodes at distance $l$ \ .}
\]
The dissimilarity between two networks is then given by the Kullback-Leibler divergence between their respective portraits.

\begin{figure}[!h]
\centering
\includegraphics[width=6in]{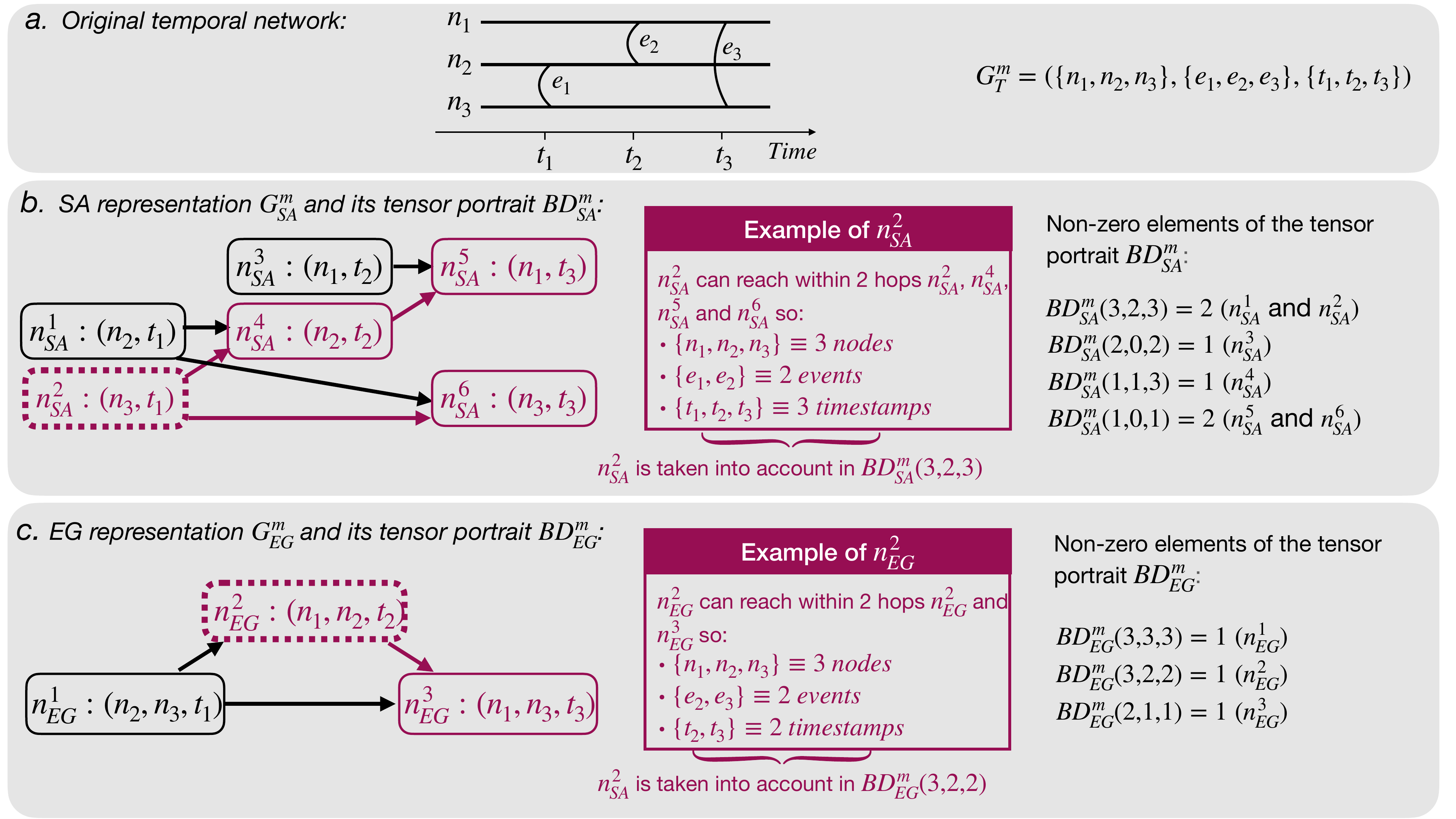}
\caption{Sketch of the method to compute the tensor portraits $BD^m_{*}$ of the temporal network $G^m_T$ displayed in panel a. The static SA and EG representations $G^m_{SA}$ and $G^m_{EG}$ are shown respectively in panels b and c. We first evaluate the number of nodes, events and timestamps from the temporal network accessible within two hops from each node of the static networks. We illustrate the computation for $n^2_{*}$ (purple highlight). We then count the number of nodes of the static representation that can reach $j$ nodes, $k$ events and $\tau$ timestamps of the original temporal network to compute the element $BD^m_{*}(j, k, \tau)$ of the tensor portrait.
}
\label{appendixb}
\end{figure}

In our case, the static networks that we need to compare are representations of temporal networks, with either the {Supra-Adjacency} or the {Event-Graph} method, noted $G_{*}^m$. Nodes and edges in these networks contain information about nodes, interactions and times of the original temporal network. To take this into account, we adapt and modify the definition of network portrait, and define the tensor portrait of $G_{*}^m$ by relying on $BD_{*}^m(j, k, \tau)$ which is the number of nodes of $G_{*}^m$ which can reach, in two hops, $j$ nodes, $k$ events and $\tau$ timestamps of the original temporal network. In other words, we consider for each node of the static representation $G_{*}^m$ its ego-network at distance $2$, and count the number of distinct nodes, timestamps and events of the original temporal network $G_T^m$ involved. We then collect this information for all nodes of $G_{*}^m$ and summarize the resulting histogram as the portrait $BD_{*}^m$. We illustrate this method to compute the tensor portrait $BD_{*}^m$ in Figure \ref{appendixb}.

We also note that the static representation of the temporal networks are directed, with edge directions following the arrow of time. The ego-network of a node of the static representation involves only future timestamps and events. To take also into account 
how each node can receive information from events in the past, we create for each
$G_{*}^m$ its reversed version $G_{*,R}^m$ by inverting the direction of each edge of the representation and compute its portrait $BR_{*}^m$. We then obtain the final tensor $B_{*}^m$ by summing $BD_{*}^m$ and $BR_{*}^m$.

Finally, we compute the dissimilarity between each pair of consecutive tensors $B_{*}^m$ and $B_{*}^{m+1}$, as their Kullback-Leibler divergence (if the two tensors $B_{*}^m$ and $B_{*}^{m+1}$ differ in size, we adjust the size of the smaller one to the size of the biggest one by filling the missing entries with zeros). The dissimilarity function is defined as:
 \[
D_{*}(m) = KL(B_{*}^m, B_{*}^{m+1}) \text{\quad for} \, m \in \mathbb{N}
\]
Note that, in the event of an empty network, the Kullback-Leibler divergence is not defined. We then assign one single event to the corresponding empty temporal sub-network $G_T^m$.

\newpage

\section{Sensitivity analysis: size and length of the temporal network, sliding window parameters}
\label{appendixa}

To evaluate the reliability of our results, we perform a sensitivity analysis on the parameters of the experiments \emph{Change of activity} and \emph{Change of grouping}. We change one parameter in both experiments while keeping the other constant ($N=100$, $\epsilon=0.001$, $\eta=45$, $\mu$ oscillating between $1.8$ and $2.8$ for the \emph{Change of activity} and $\mu=2.8$ for \emph{Change of grouping}, $T_a=T_g=100$, $\mid T \mid$ is adjusted to have 12 periods, $\Delta t_w=5$ and $t_w=2$). 

\begin{figure}[!h]
\centering
\includegraphics[width=6in]{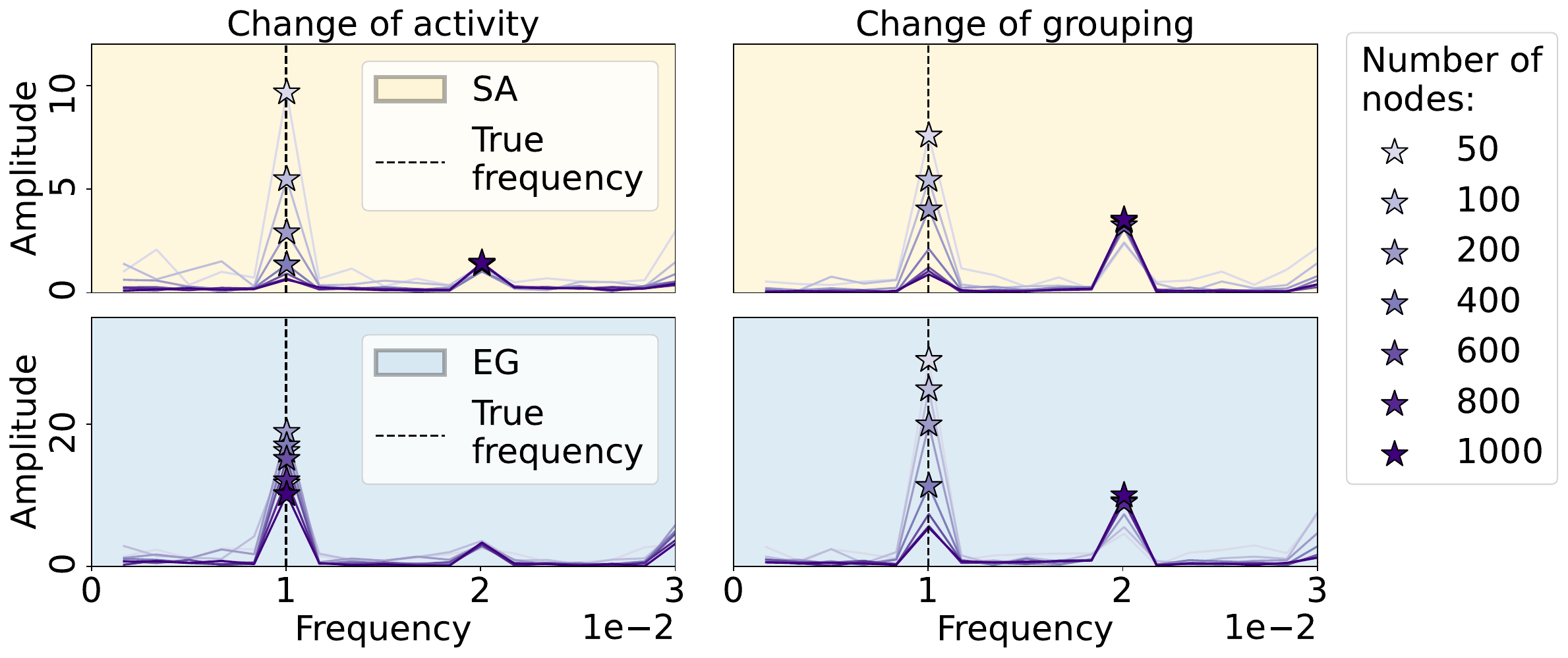}
\caption{Fourier transforms of the temporal network from the \emph{Change of activity} experiment (left column) and the \emph{Change of grouping} experiment (right column) measured from the {SA-method} (yellow background) and the {EG-method} (blue background). The number of nodes of the AD networks varies from 50 to 1000. The correct frequencies are indicated with vertical dashed lines. Those original periods are well-measured in the majority of the cases. However, when the number of nodes is too important, the method measures the semi-period.}
\label{appendix1a_1}
\end{figure}

When we vary the number of nodes of the networks (Figure \ref{appendix1a_1}), the original period is almost always measured properly except when the number of nodes is high. In that case, the measured period is $50$ (corresponding to a frequency of $0.02$), which is half of the original period. In fact, when the networks change from a high-activity state to a low-activity state, we observe a peak in the dissimilarity function. This situation happens twice in a period: when changing from low to high and from high to low activity. The measured period is then the half period (Note that this happens at all sizes, and the half-period is indeed always recovered as one of the harmonics, but it seems here to become dominant at large sizes). 

\begin{figure}[!h]
\centering
\includegraphics[width=6in]{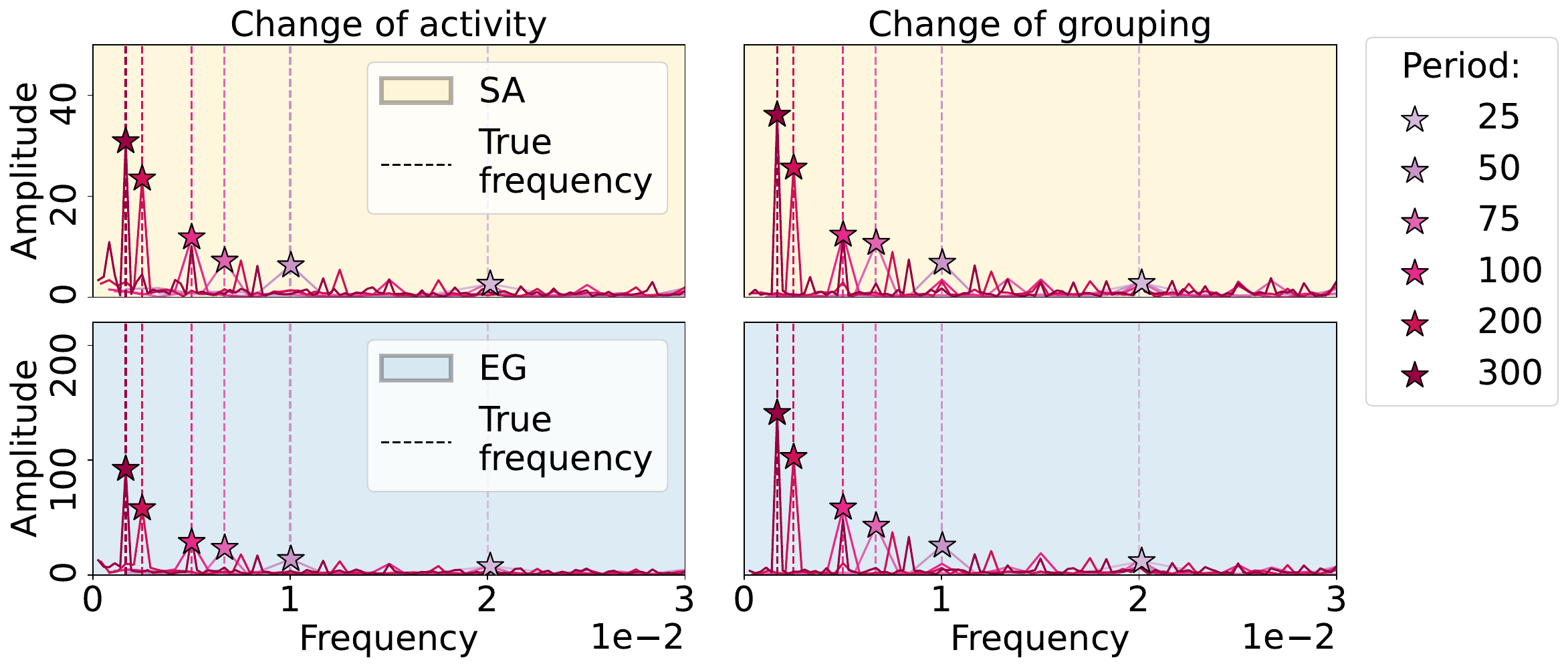}
\caption{Fourier transforms of the temporal network from the \emph{Change of activity} experiment (left column) and the \emph{Change of grouping} experiment (right column) measured from the {SA-method} (yellow background) and the {EG-method} (blue background). The period varies from $25$ to $300$. The correct frequencies are indicated with vertical dashed lines and are here always well recovered.}
\label{appendix1a_2}
\end{figure}

\begin{figure}[!h]
\centering
\includegraphics[width=6in]{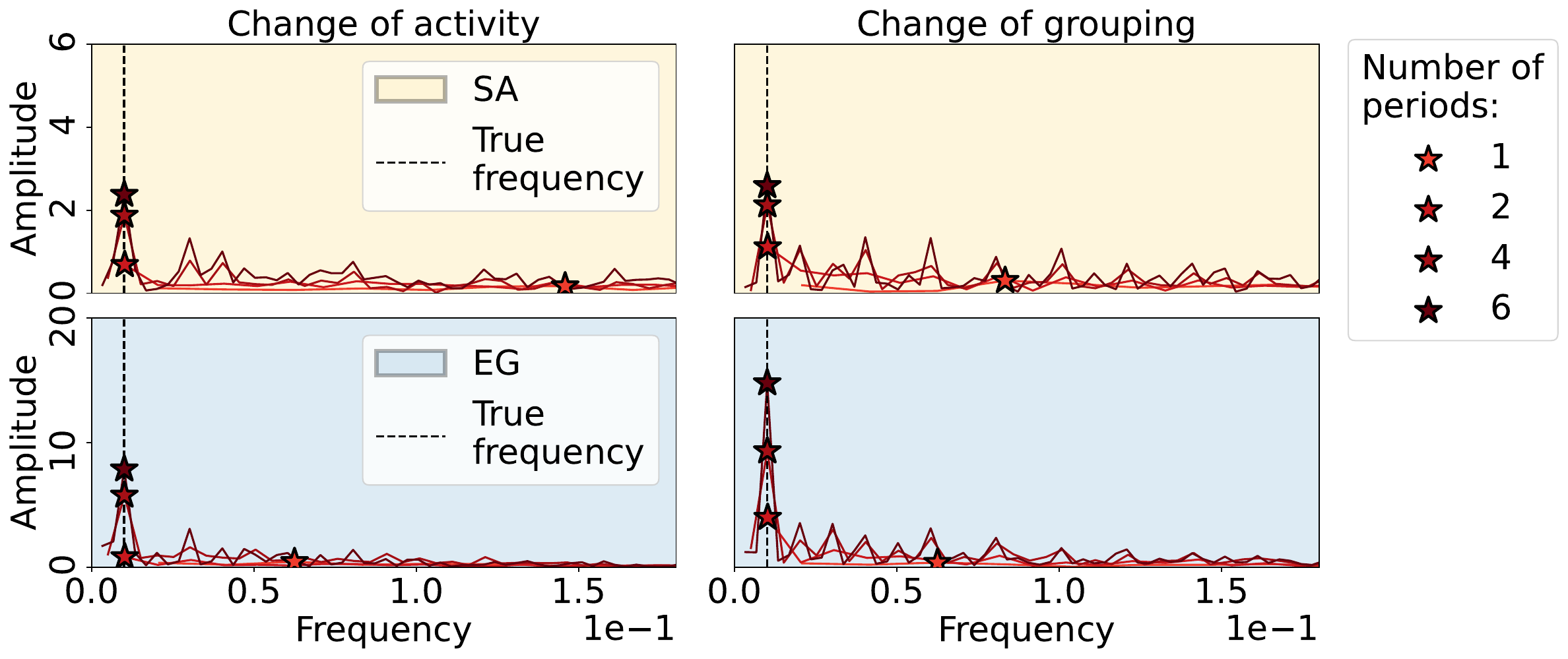}
\caption{Fourier transforms of the temporal network from the \emph{Change of activity} experiment (left column) and the \emph{Change of grouping} experiment (right column) measured from the {SA-method} (yellow background) and the {EG-method} (blue background). The number of periods of the AD networks observed during $T$ varies from 1 to 6. The correct frequencies are indicated with vertical dashed lines. Those proper periods are well-measured as long as the data set contains at least two periods.}
\label{appendix1a_3}
\end{figure}

The same analysis has been realized by changing the length of the period (Figure \ref{appendix1a_2}) and in every case, the original period is correctly measured: the length of the period does not influence the observation. Instead, when we change the number of periods observed (Figure \ref{appendix1a_3}), we observe that we need a minimum of 2 periods to measure the original time scale.

\begin{figure}[!h]
\centering
\includegraphics[width=0.9\textwidth]{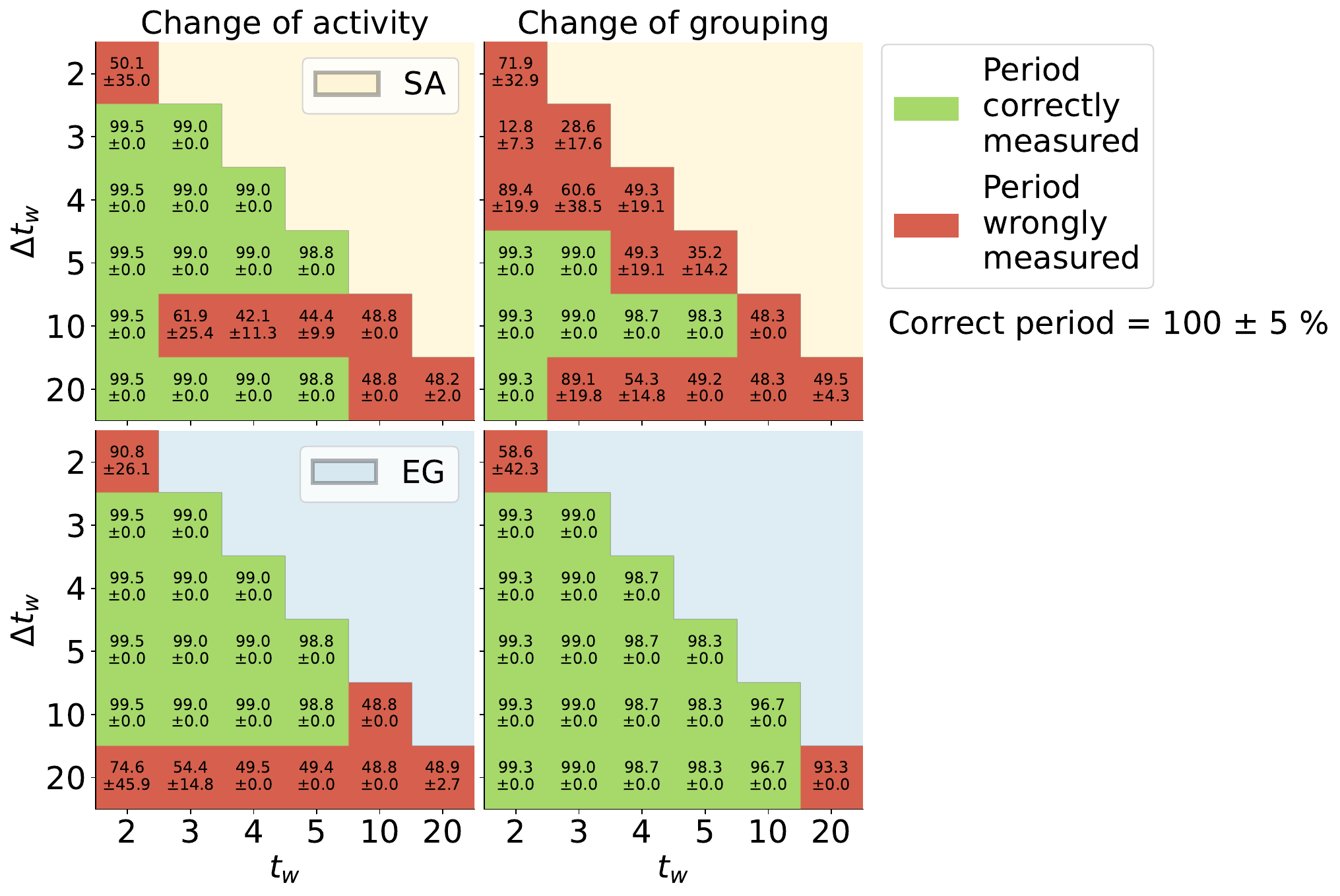}
\caption{Period measured through the {SA-method} (orange background) and the {EG-method} (blue background) for the \emph{Change of activity} experiment (first column) and the \emph{Change of grouping} experiment (second column). We change the parameters of the sliding window: the x-axis presents different values of the stride $t_w$ and the y-axis different values of time-window lengths ($\Delta t_w$). The period of the initial networks is 100 and the results are averaged over 10 realisations. The results are displayed as the average period over the different realisations $\pm$ the standard deviation. 
}
\label{appendix1b}
\end{figure}

We finally study the influence of the parameters of the sliding window $t_w$ and $\Delta t_w$ on the results. In Figure \ref{appendix1b}, we compute the Fourier Transform of the AD network with the {SA-method} and the {EG-method} having different parameters $t_w$ and $\Delta t_w$. The correct period ($100$) is properly measured if $t_w$ and $\Delta t_w$ are not too large.

\clearpage
\newpage

\section{Empirical data}
\label{appendixc}

\begin{figure}[!h]
\centering
\includegraphics[width=6in]{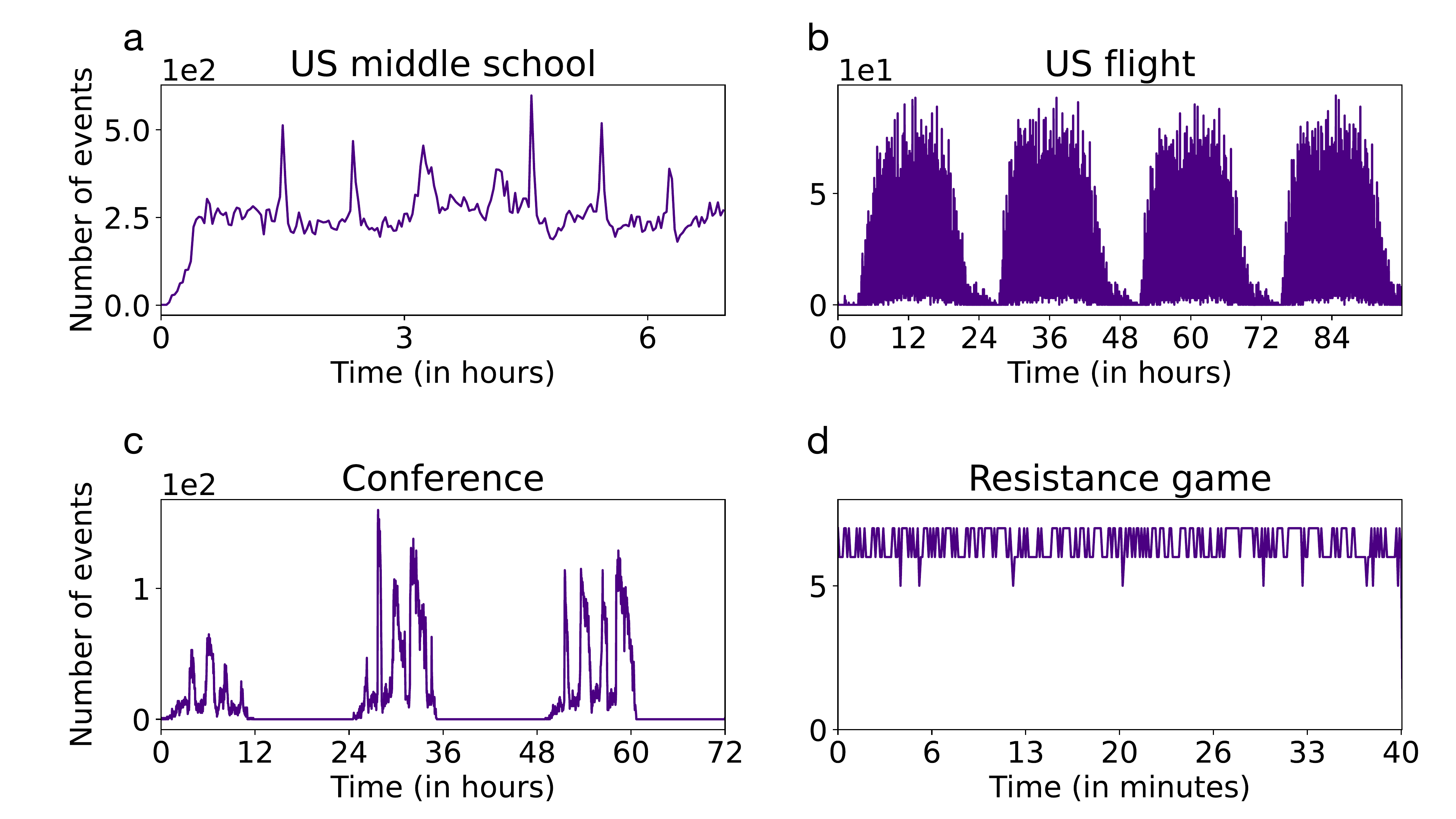}
\caption{Number of events as a function of time for the four data sets: the \emph{US school} (panel a), the \emph{US flight} (panel b), the \emph{Conference} (panel c) and the \emph{Resistance game} (panel d). The \emph{US school} network contains high activity periods during recreational moments of the students' day, while the \emph{US flight} and the \emph{Conference} networks present circadian patterns. The {Resistance game} network does not have particular periodic activity changes.
}
\label{fig_sim}
\end{figure}

We observe the changes of activity over time of the empirical networks for the four data sets (Figure \ref{fig_sim}). The \emph{US school} presents periodic patterns, varying from low contact periods when the students are in class to high contact periods when there is a recreational time. The \emph{US flight} and \emph{Conference} networks have circadian patterns as there are respectively less flights and less contacts at night. Finally, the \emph{Resistance game} does not present any periodic change in its activity as every player of the game is looking at someone else at each time step.

\end{document}